\documentclass[%
 reprint,
 amsmath,amssymb,
 aps,
]{revtex4-2}

\usepackage{graphicx}
\usepackage{dcolumn}
\usepackage{bm}
\usepackage{mathrsfs}
\usepackage{float}      
\usepackage[table]{xcolor}      
\usepackage{subcaption} 
  
\usepackage{multirow}
\usepackage{makecell}
 \usepackage{amsmath}
\usepackage{graphicx}
\usepackage{url}
\usepackage{fancyhdr}
\usepackage{footmisc}
\usepackage{algorithmic}
\usepackage{listings}
\usepackage{fancyvrb}
\usepackage{graphicx}
\usepackage{hyperref}
 
\begin{document}

\graphicspath{ {./} }
\preprint{APS/123-QED}
\title{Symmetry considerations in exact diagonalization: spin-1/2 pyrochlore magnets}

\author{C.~Wei}
\email{cw7734@mun.ca}
\author{S.~H.~Curnoe}%
 \email{curnoe@mun.ca}
\affiliation{%
 Department of Physics and Physical Oceanography, Memorial University of Newfoundland,
St. John's, Newfoundland \& Labrador, Canada A1B 3X7
}%

\date{\today}

\begin{abstract}
We describe how the methods of group theory (symmetry) are used to optimize the problem of exact diagonalization of a quantum system on a 16-site pyrochlore lattice. By analytically constructing a complete set of symmetrized states, we completely block-diagonalize the Hamiltonian. As an example, we consider a spin-1/2 system with nearest neighbour exchange interactions. 
\end{abstract}

                              
\maketitle

\section{Introduction}\label{sec:Intro}

Exact diagonalization uses numerical approaches to find the eigenvalues and eigenvectors of a matrix representation of a Hamiltonian. The results determine the entire spectrum, which can be used to evaluate many quantities of interest, including spin correlations, thermodynamic quantities, and even quantum entanglement. Generally though, these studies are limited by size, since the dimension of the Hamiltonian matrix is $D\times D$, where for $N$ spin-1/2 states $D=2^N$, the memory requirement scales as $D^2$, and the computational time scales as $D^3$ \cite{Zhang_2010}. 
Different strategies can be used to circumvent the size problem. To begin with, it is often possible to block-diagonalize the Hamiltonian matrix by using the underlying symmetries of the system. 
Then, if the full spectrum is not needed, the Lanczos algorithm may used to solve low-lying eigenvalues and eigenstates \cite{lanczos_1950,wei_Alexander_and_Fehske_H}, and has been applied to general pairing Hamiltonians of size $10^{8-9}$.
With space group  $Fd\bar{3}m$ ($O_h^7$, No.~227), pyrochlore crystals are highly symmetric; in
this article we describe the procedure for block-diagonalizing the spin Hamiltonian for pyrochlore crystals by fully exploiting 
the space group symmetries. 
 


Pyrochlore magnets have been a popular research subject for the past few decades because they are physical realizations of spin systems on a geometrically frustrated lattice. In these crystals, the magnetic ions with spin $J$ reside on the vertices of a network of corner-sharing tetrahedra. 
The crystal electric field at the magnetic sites lifts the $2J+1$-fold spin degeneracy into singlets and doublets; those with a well-separated ground state doublet are effective spin-1/2 systems.  
The simplest model of magnetic interactions in pyrochlore crystals is the nearest-neighbour exchange interaction, which accounts for 
the interaction energy between pairs of neighbouring spins. Given the non-isotropic electronic structure of the magnetic ions (which are typically rare earth elements) the magnetic interaction is expected to be anisotropic ({\em i.e.} non-Heisenberg) in general. Even so, 
this model is tightly constrained by the space group symmetry of the crystal such that the general form of the nearest-neighbour exchange interaction has only four free parameters \cite{Curnoe_physrevb_2008};
this same symmetry group is used to block-diagonalize the Hamiltonian matrix.  

Pyrochlore magnets exhibit a variety of magnetic phenomenon, including various magnetically ordered states and different kinds of `spin ice,' such as disordered spin ice in Ho$_2$Ti$_2$O$_7$\cite{Ho2Ti2O7}, ordered spin ice in Tb$_2$Sn$_2$O$_7$\cite{Tb2Sn2O7} and quantum spin ice in Tb$_2$Ti$_2$O$_7$\cite{Gardne_physrevb_2001, curnoe_2013_PhysRevB.88.014429}, Yb$_2$Ti$_2$O$_7$ \cite{Ross_physrevX_2011,chang_2021_nature_commun, pan_nature_2014}, Ce$_2$Zr$_2$O$_7$ \cite{gaudet_2019_PhysRevLett.122.187201}.  Quantum spin ice materials in particular have attracted a great deal of attention because they are believed to host long-range quantum entanglement. In a recent work, we have studied the Hamiltonian within a range of its four free parameters that encompasses this state \cite{chen_curnoe_2023}; here we provide the details of our computational method.

\section{Exact Diagonalization}\label{sec:Model}

\subsection{Representations of the 
Symmetry Group}
 The symmetry group of any crystal is one of the 236 crystallographic space groups \cite{international_TableA}, each of which consists of a set of translations based on the crystal lattice vectors and a point group of rotations. All of the elements of the space group can be represented as matrices. An {\em irreducible representation} (IR) is any set of matrices that {\em cannot} be block-diagonalized by a unitary transformation; reducible representations are a direct sum of IR's.  Explicit forms for the matrices can be found using the methods described in Ref.\ \cite{Kovalev_book_1965}.

A finite system with periodic boundary condition will support a number of wavevectors $k$ equivalent to the size of the translation subgroup. 
To each wavevector $k$ is associated one or more IRs, depending on 
the symmetry group of $k$ itself.   For pyrochlore crystals, which have a face-centred cubic (fcc) Bravais lattice, the only translations in the symmetry group of a single cube are the set of 
 three fcc lattice vectors
 $\vec{\tau}_1 = (0,a/2,a/2)$, 
 $\vec{\tau}_2 = (a/2,0,a/2)$ and
 $\vec{\tau}_3 = (a/2,a/2,0)$ 
 (where $a$ is the side length of the cube) and in $k$-space only the 
 $\Gamma$-point ($\vec{k} = (0,0,0))$
 and the $X$-point
 ($\vec{k} = \frac{2\pi}{a}(0, 0, 1)$ and equivalent) occur. A system of two cubes, with 
 periodic boundary conditions
 $f(x,y,z) = f(x\pm a,y\pm a,z) =
 f(x\pm a,y,z\pm a) = 
 f(x,y\pm a, z\pm a)$, has four additional translations, and in $k$-space, in addition to the $\Gamma$-point and the $X$-point, the $L$-point,
 with $\vec{k} = \frac{\pi}{a}(1,1,1)$, occurs.  Altogether there are ten $\Gamma$-point IRs (four one-dimensional, two two-dimensional and four three-dimensional),
 four $X$-point IRs (all
 six-dimensional) and six $L$-point IRs (four four-dimensional and two eight-dimensional). The dimensionality of the IR is its degeneracy. This means that 
 the Hamiltonian matrix for
 a cube (containing 16 magnetic sites)
 can be block-diagonalized into 
 44 blocks, of which only fourteen are not redundant, which represents a computation time reduction by a factor of approximately 6000.

 A matrix representation of the symmetry group can be found by considering the action of the symmetry group on a set physical objects, such as the set of basis kets $|u_i\rangle$ of the system. 
 Using this basis, the matrix elements for a symmetry operation $R$ are $\Gamma_{ij}(R) = \langle u_i|R|u_j\rangle$,
 $R|u_i\rangle = \sum_{j} \Gamma_{ij}(R)
 |u_j\rangle$. 
 Generally such constructions will generate a reducible representation 
 which can be decomposed into
IR's $\Gamma^{(i)}$ \cite{Tinkham_book_1992}:
 \begin{equation}
     \Gamma =  \sum_{\oplus i} a_i\Gamma^{(i)}.
 \end{equation}
where $a_i$ (a non-negative integer) is the number of copies of the $i$th
IR.  
Since they are invariant under a unitary transformation, comparing the character (traces) of $\Gamma$ with the characters of the $\Gamma^{(i)}$ allows one to extract the numbers $a_i$. 
By applying a suitable unitary transformation, the matrices of $\Gamma$ can be cast into a block-diagonal form where the blocks are of size $a_i$.  
If the same basis is used to 
generate a matrix representation of the Hamiltonian, with matrix elements
$H_{ij} = \langle u_i|H|u_j\rangle$, then $H$ can be block-diagonalized by the same unitary transformation.
In the following, we describe how to construct the unitary matrix that block-diagonalizes the Hamiltonian for pyrochlore magnets.

\subsection{Application to Pyrochlore Magnets}

In a spin-1/2 system with $N$
spins there are $2^N$ basis kets of the form
$|\pm\pm\pm ...\rangle$ where the order of the symbols inside the ket corresponds to 
some particular order of the spin sites on the lattice. We now describe how these kets form the basis of a reducible representation of
the symmetry group of pyrochlore crystals.




In pyrochlore magnets, the magnetic ions occupy the 16d Wyckoff position, such that there are 16 ions in the cubic cell.  The primitive unit cell is a tetrahedron, with magnetic ions located at each of its four vertices; this is the smallest unit that possesses the full point group symmetry of the pyrochlore crystal, $O_h$. The $2^4\times 2^4$ single-tetrahedron problem is easily solved without the need 
for block-diagonalization, and block-diagonalization renders the problem almost trivial.  The set of $2^4$ basis kets $|\pm\pm\pm\pm\rangle$  belong to 
the reducible representation
$A_{1g} \oplus 3E_g \oplus 2 T_{1g} \oplus
T_{2g}$, where $A$ is one-dimensional, $E$ is two-dimensional, and $T$ is three-dimensional, and all belong to the $\Gamma$-point. Thus the $16\times 16$ problem is
reducible to blocks of size $1\times 1$ (non-degenerate), size $3 \times 3$ (doubly degenerate), size $2\times 2$ (triply degenerate), and size $1\times 1$ (triply degenerate) \cite{Curnoe2007_PhysRevB_2007}.

\begin{widetext}
\begin{center}
\begin{table}[h]
\begin{tabular}{|r|r|r|r|r|r|r|r|r|r|r|r|r|r|r|}
\hline
	&$1E$&	$6C_2$	&$32C_3$&	$12C_2'$&	$24C_4$&	$4I$&	$12IC_2$&	$32IC_3$&	$12IC_2'$&	$12IC_2'$&	$24IC_4$&	$3\tau$&	$6\tau C_2$&	$12\tau C_2'$
   \\ \hline
$A_{1g}$&	1&	1&	1&	1&	1&	1&	1&	1&	1&	1&	1&	1&	1&	1
  \\ \hline
$A_{2g}$&	1&	1&	1&	-1&	-1&	1&	1&	1&	-1&	-1&	-1&	1&	1&	-1
 \\ \hline
$A_{1u}$&	1&	1&	1&	1&	1&	-1&	-1&	-1&	-1&	-1&	-1&	1&	1&	1
 \\ \hline
$A_{2u}$&	1&	1&	1&	-1&	-1&	-1&	-1&	-1&	1&	1&	1&	1&	1&	-1
\\ \hline
$E_{g}$&	2&	2&	-1&	0&	0&	2&	2&	-1&	0&	0&	0&	2&	2&	0
   \\ \hline
$E_{u}$&	2&	2&	-1&	0&	0&	-2&	-2&	1&	0&	0&	0&	2&	2&	0
  \\ \hline
  $T_{1g}$&	3&	-1&	0&	-1&	1&	3&	-1&	0&	-1&	-1&	1&	3&	-1&	-1
 \\ \hline
$T_{2g}$&	3&	-1&	0&	1&	-1&	3&	-1&	0&	1&	1&	-1&	3&	-1&	1
\\ \hline
$T_{1u}$&	3&	-1&	0&	-1&	1&	-3&	1&	0&	1&	1&	-1&	3&	-1&	-1
 \\ \hline
$T_{2u}$&	3&	-1&	0&	1&	-1&	-3&	1&	0&	-1&	-1&	1&	3&	-1&	1
  \\ \hline

$X_1$&	6&	2&  0&	2&	0&	0&	0&	0&	0&	0&	0&	-2&	-2&	-2
    \\ \hline
$X_2$&	6&	2&	0&	-2&	0&	0&	0&	0&	0&	0&	0&	-2&	-2&	2
      \\\hline
$X_3$&	6&	-2&	0&	0&	0&	0&	0&	0&	-2&	2&	0&	-2&	2&	0
      \\ \hline
$X_4$&	6&	-2&	0&	0&	0&	0&	0&	0&	2&	-2&	0&	-2&	2&	0
       \\ \hline
\end{tabular}
   \caption{Character table for the symmetry group of a pyrochlore crystal cube containing $16$ magnetic sites, $\{O_h\}\times \{E,\tau_1, \tau_2, \tau_3\}$.
   The first row lists the classes of the group (the number of elements in each class and a representative element in the class: $C_2$ is a 180$^\circ$ rotation about a principle cube axis, $[100]$; $C_3$ is a 3-fold rotation about the cubic diagonal, $[111]$; $C_2'$ is a 180$^\circ$ rotation about the $[110]$ axes; $C_4$ is a 4-fold rotation about a principle cube axis; $I$ is inversion; and $\tau$ is an fcc lattice translation).
   The first column lists the irreducible representations and the second column lists their dimensions.   The character of each representation is the set of numbers in each row, which are the traces of the matrices representing the group operations.
   \label{16site-char} } 
\end{table} 
 \end{center}
\end{widetext}

\begin{figure}
\centering
  \includegraphics[width=80mm]{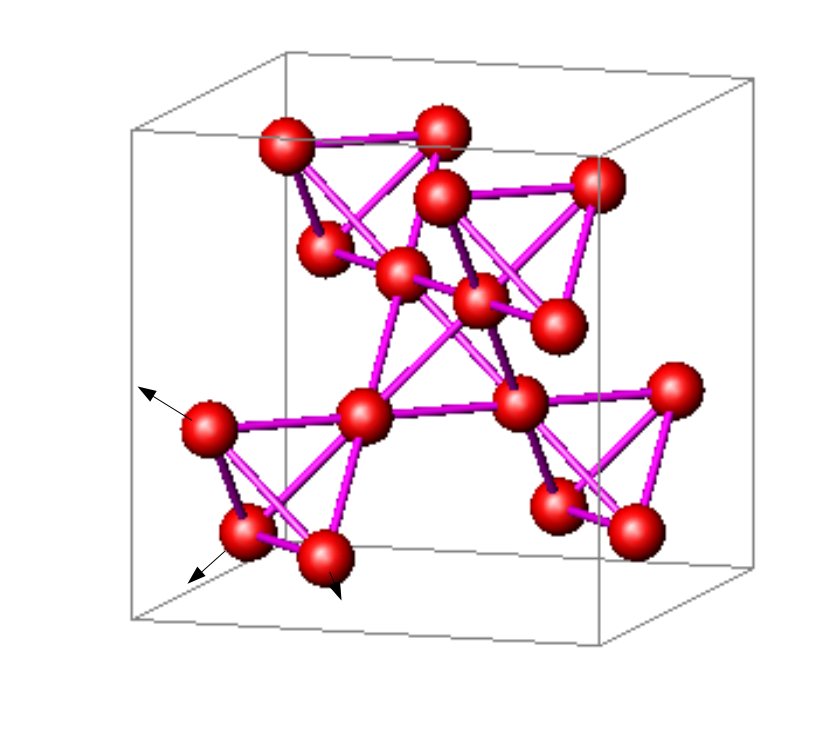}
\caption{Magnetic sites of a pyrochlore crystal. The arrows are axes of 3-fold symmetry.\label{fig:pyro}}
\end{figure}

The next largest symmetric unit is a cube containing 16 magnetic ions (shown in Fig.\ \ref{fig:pyro}), which is the focus of this article.
With periodic boundary conditions assumed, there are only three translations, $\Vec{\tau}_1$, $\Vec{\tau}_2$ and $\Vec{\tau}_3$. The symmetry group is $\{O_h\}\times \{E, \tau_1, \tau_2, \tau_3\}$, which has $192$ elements.  The character table of this group is given in Table \ref{16site-char}.  The decomposition of the representation generated by the set of basis kets 
$|\pm\pm\pm\pm\pm\pm\pm\pm\pm\pm\pm\pm\pm\pm\pm\pm\rangle$ is given in Table
\ref{16blocks}.  According to these results, the matrix representation of any operator 
that that acts on this basis and is invariant under this symmetry group can be block-diagonalized into 44 blocks with different sizes and degeneracies, as listed in Table~\ref{16blocks}.

\begin{table}[h]
\begin{tabular}{c|c|c } 
\hline 
IR & dimension & block size 
\\
\hline 
{A$_{1g}$} & 1& 383\\ 
{A$_{2g}$} & 1&371 \\
{A$_{1u}$}& 1&335 \\
{A$_{2u}$} & 1&335 \\
{E$_{g}$} & 2&774\\
{E$_{u}$} & 2&682\\
{T$_{1g}$}& 3&1085 \\
{T$_{2g}$}& 3&1081 \\
{T$_{1u}$} & 3&957 \\
{T$_{2u}$} & 3&957 \\
{X$_{1}$} & 6&2038 \\
{X$_{2}$}& 6&2042 \\
{X$_{3}$} & 6&2038 \\
{X$_{4}$} & 6&2042 \\
\hline 
\end{tabular}
   \caption{The decomposition of the representation generated by the $2^{16}$ basis kets of a cube containing 16 sites. 
    The first column lists the IR's of the 
   symmetry group, the second column lists their dimension (degeneracy), and the final column gives the size of each block (the number of copies $a_i$ of each IR).
    \label{16blocks}}
\end{table}

As an application of our method, we consider the general nearest-neighbor exchange interaction for pyrochlore magnets. The Hamiltonian is
\begin{equation}
    H_{ex}=\sum_{\langle i,j\rangle}{\cal J}_{i,j}^{\mu \nu}S^\mu_i S^\nu_j,
\end{equation}
where the sum over $\langle i,j\rangle$ runs over pairs of nearest-neighbour spins
and $\vec{S}_i=(S^x_i, S^y_i,S^z_i)$ is the spin operator for the $i$th site. 
${\cal J}_{i,j}^{\mu \nu}$ are exchange constants which are constrained by the space group symmetry of the crystal; in pyrochlore magnets
there are only four independent exchange constants. 
It is convenient to express the Hamiltonian as
\begin{equation}
H = {\cal J}_{1}X_1+{\cal J}_{2}X_2+{\cal J}_{3}X_3+{\cal J}_{4}X_4,
\label{eq:hamil}
\end{equation}
where ${\cal J}_{a}$ are the exchange constants and
\begin{align*}
&X_1=-\frac{1}{3}\sum_{\langle i,j \rangle}S_{iz}S_{jz}\\
&X_2=-\frac{\sqrt{2}}{3}\sum_{\langle i,j\rangle}[\Lambda_{s_is_j}(S_{iz}S_{j+}+S_{jz}S_{i+})+{\rm h.c.}]\\
&X_3=\frac{1}{3}\sum_{\langle i,j\rangle}[\Lambda_{s_is_j}^*S_{i+}S_{j+}+ {\rm h.c.}]\\
&X_4=-\frac{1}{6}\sum_{\langle i,j \rangle}(S_{i+}S_{j-}+ {\rm h.c.}).
\end{align*}
$S_{\pm} = S_{x} \pm i S_y$ and the subscript is
used to indicate the components of the spin operators with respect to a set of local axes, defined as follows. The local symmetry of the magnetic sites (the 16d Wyckoff position) is the point group $D_{3d}$, which has a three-fold axis pointing along one of the four cube diagonals; the local $z$-axis is defined to be this three-fold axis (see Fig.\ \ref{fig:pyro} and Refs.\ \cite{Curnoe2007_PhysRevB_2007, Curnoe_physrevb_2008} for
more details). 
$\Lambda_{ss'}$ are phases which depend on the site numbers: $\Lambda_{12}=\Lambda_{34}=1$ and $\Lambda_{13}=\Lambda_{24}=\Lambda_{14}^{*}=\Lambda_{23}^{*}=\varepsilon\equiv \exp(\frac{2\pi i}{3})$. 
The Hamiltonian (\ref{eq:hamil}) is the most general form allowed by symmetry for nearest-neighbour interactions with angular momentum operators of any value of $S$ (including the classical limit where $S$ is large), however, in the following, we assume that $\vec{S}$ is a spin-1/2 operator. One may also construct general Hamiltonians very similar in form  for `pseudo-spin' operators, where magnetic sites are occupied by ions 
with a double-degeneracy that is 
different from spin-1/2, \cite{curnoe_condensedmatter_2018} but we will not consider those models here.

The basis kets $|u_i\rangle = |\pm\pm\pm \ldots \rangle$ represent states where
the quantization axes of the spins are the local $z$-axes described above. Hence the action of the spin operator is 
\begin{eqnarray}
S_{iz}|\ldots \pm \ldots \rangle &=& \pm\frac{\hbar}{2}|\ldots \pm \ldots\rangle \\
S_{ix}|\ldots\pm\ldots\rangle &= & \frac{\hbar}{2}|\ldots\mp\ldots\rangle \\
S_{iy}|\ldots\pm\ldots\rangle &=&\pm\frac{i\hbar}{2}|\ldots\mp\ldots\rangle .
\end{eqnarray}
Since the basis kets are eigenstates of $S_{iz}$, the matrix representation of the term $X_1$ will be a diagonal  in the $|u_i\rangle$ basis. However,  $X_2$, $X_3$ and $X_4$ will be non-diagonal (albeit sparse).  

To find the unitary matrix $U$ that block-diagonalizes $H$, we find the set
of ``symmetrized" kets, each of which belongs to a particular IR, as follows. We first obtain explicit matrices for  
all of the group operations in all of the IR's following the procedure in Ref.\ \cite{Kovalev_book_1965}. 
In the notation of Ref.\ \cite{Tinkham_book_1992},
the matrix elements of the group operations are
$\Gamma^{(j)}_{\lambda\kappa}(R)$ where $j$ labels
the representation, $R$ is an element of the symmetry group, and 
$\lambda$ and $\kappa$ are the row and column of the matrix.  
We also find the action of every symmetry element $R$ on every basis ket.  Essentially, every $R$ produces a permutation of the basis kets and may introduce phases.
The operator
$$
{\cal P}_{\kappa\kappa}^{(j)} = \sum_R \Gamma^{(j)*}_{\kappa\kappa} P_R,
$$ 
where $P_R$ is the is the operator that applies the 
symmetry element $R$ to a ket, is proportional to a projection operator for
the $\kappa$th row of 
the $j$th IR.  When this operator is applied to a basis ket $|u_i\rangle = |\pm\pm\pm \ldots\rangle$ any non-zero result will be a symmetrized ket belonging to the 
$\kappa$th dimension of 
the $j$th IR.  The brute force strategy is to apply the projector 
to each basis ket until all $a_j$ independent symmetrized kets are found. However, there are much faster ways to implement this in practice, as discussed in Appendix B. 
The final result will be a set of symmetrized kets expressed as
$|\phi_n\rangle = \sum_i \alpha_{ni} |u_i\rangle$; these should be grouped according to the IR and its dimension ($j$, $\kappa$) to which they belong. 
The matrix elements of the unitary matrix $U$ that diagonalizes $H$,
$H_{\rm block} = U^{\dagger}HU$
are $U_{lj} = \langle \phi_l|u_j\rangle
= \alpha^{*}_{lj}$, that is
\begin{eqnarray}
H_{ij}^{\rm block} & = & 
\langle \phi_i|H|\phi_j\rangle 
 =  \sum_{kj} \langle \phi_i|u_k\rangle \langle u_k |H|u_l\rangle \langle u_k |\phi_j\rangle
 \nonumber 
\\
& = & \sum_{kl} U^{\dagger}_{ik}H_{kl} U_{lj}.
\end{eqnarray}
Finally, the eigenvalues and eigenvectors of
$H^{\rm block}$ are calculated numerically using Lapack subroutines. 


\section{Discussion and Summary}

\begin{figure}
\centering
\subfloat[${\cal J}_{2}=-0.02$ ]{  \includegraphics[width=40mm]{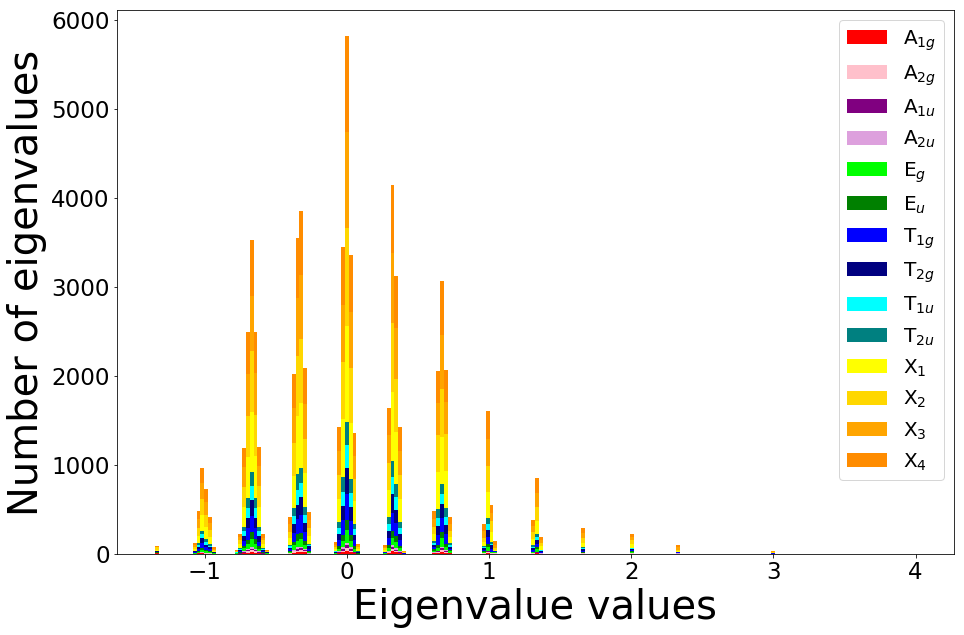}
}
\subfloat[${\cal J}_{2}=-0.08$]{
\includegraphics[width=40mm]{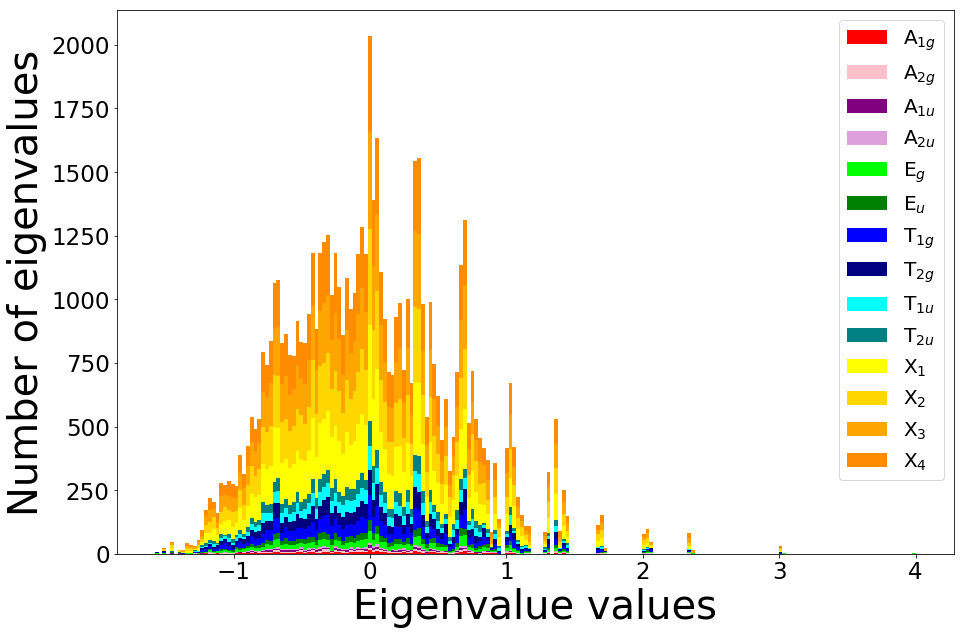}
}
\hspace{0mm}
\subfloat[${\cal J}_{2}=-0.14$ ]{ \includegraphics[width=40mm]{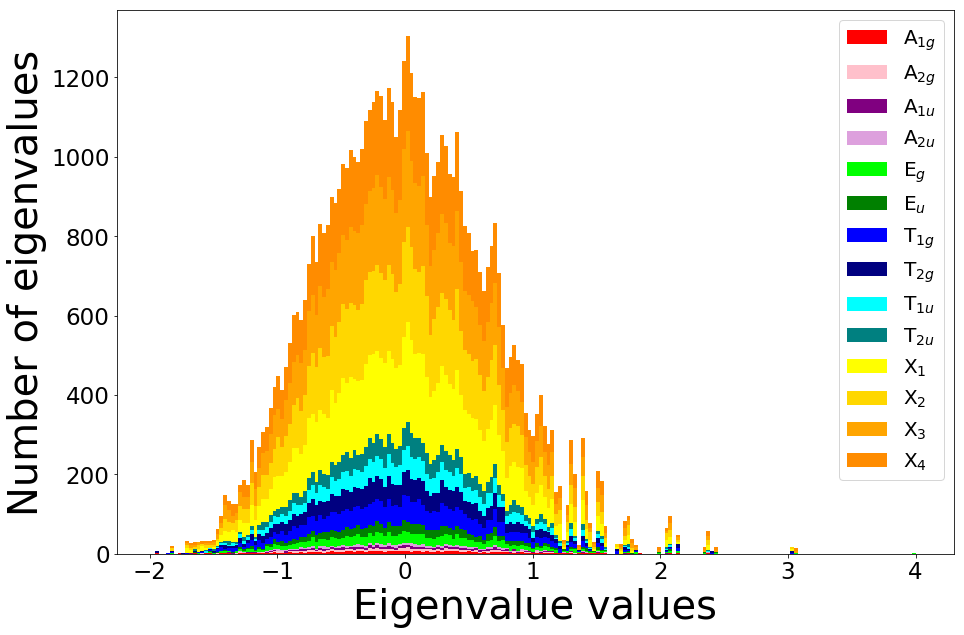}
}
\subfloat[${\cal J}_{2}=-0.2$]{  \includegraphics[width=40mm]{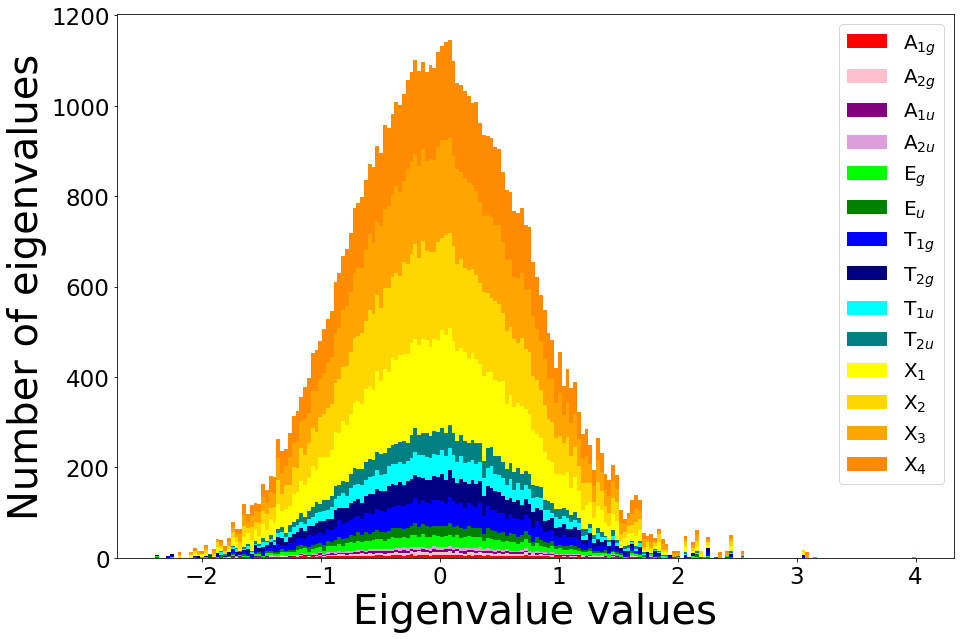}
}
\caption{Eigenvalues of $H$ 
for ${\cal J}_1 = -1$, ${\cal J}_3 = {\cal J}_4 = 0$, and  selected values of ${\cal J}_{2}$.
\label{j2_all_1}}
\end{figure}

In Figs.\ \ref{j2_all_1}-\ref{j4n_all} we present the entire spectrum of the Hamiltonian in the spin ice regime (${\cal J}_1 = -1$) for a $N=16$ site system.  
In Fig.\ \ref{j2_all_1}, the histograms are colour-coded to indicate the eigenvalues of individual blocks of the block-diagonalized $H$, labelled by IR. It is evident that each block contains the entire range of the spectrum. 

\begin{figure}[ht]
\centering
\subfloat[${\cal J}_{2}=-0.02$ ]{
  \includegraphics[width=40mm]{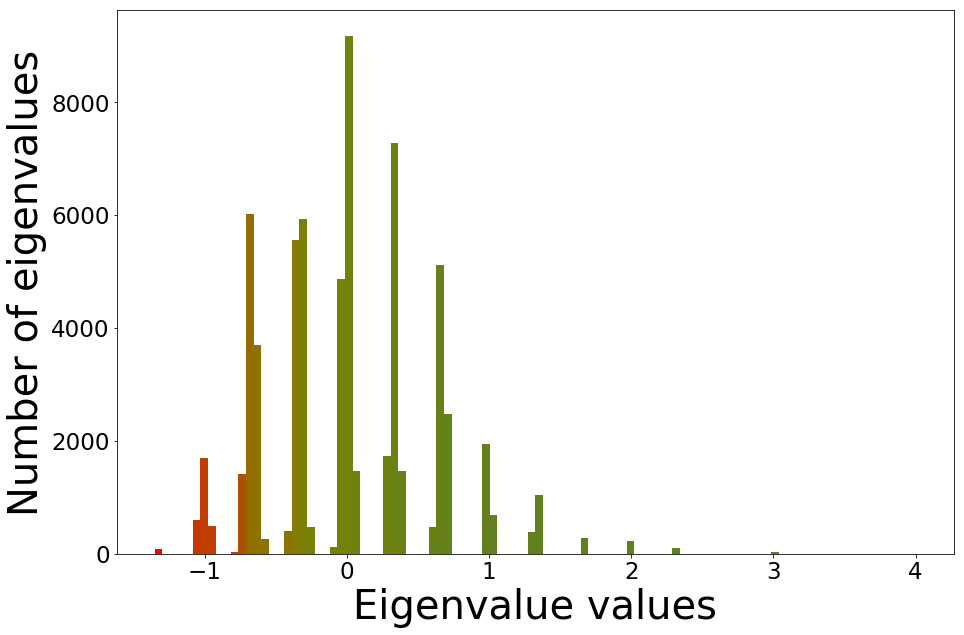}
}
\subfloat[${\cal J}_{2}=-0.08$]{
  \includegraphics[width=40mm]{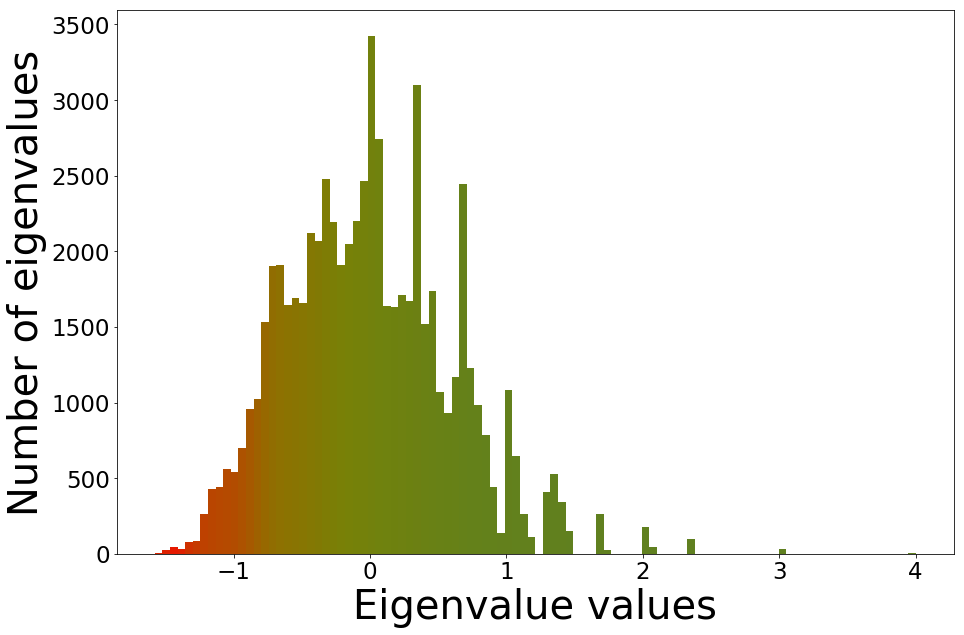}
}
\hspace{0mm}
\subfloat[${\cal J}_{2}=-0.14$ ]{
  \includegraphics[width=40mm]{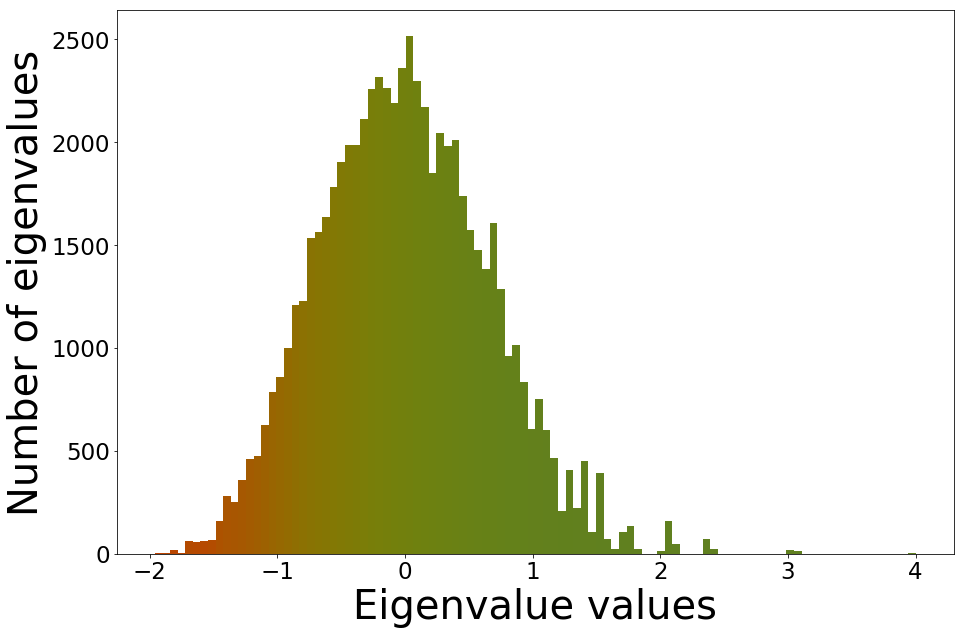}
}
\subfloat[${\cal J}_{2}=-0.2$]{
  \includegraphics[width=40mm]{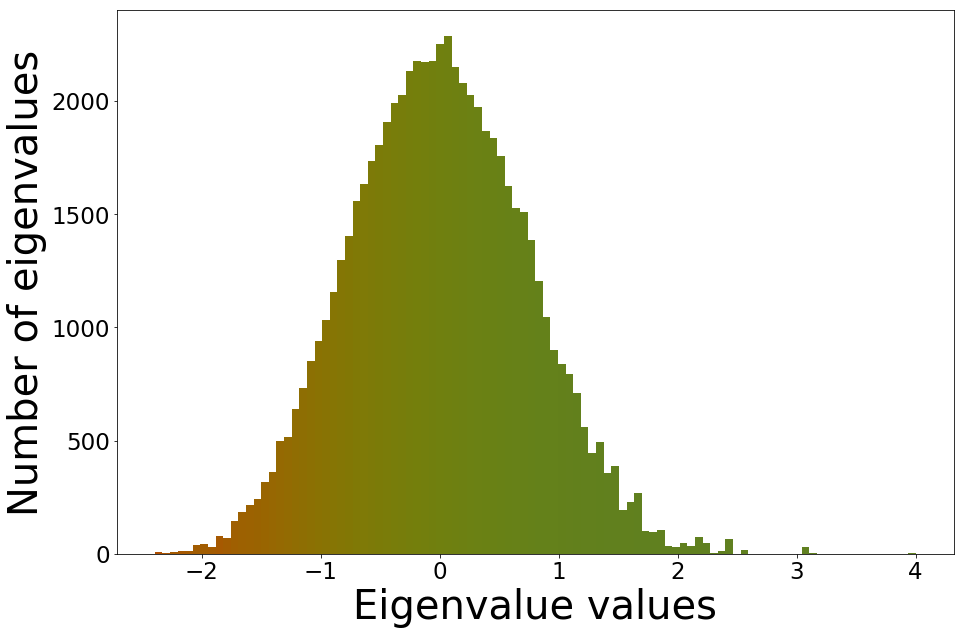}
}
\caption{Spectrum of $H$ 
for ${\cal J}_1 = -1$, ${\cal J}_3 = {\cal J}_4 = 0$,   and  different values of ${\cal J}_{2}$.
\label{j2_all}}
\end{figure}

For ${\cal J}_{2,3,4}=0$
the eigenvalues are discrete and evenly spaced,  
$E_n = n{\cal J}_1/3$ for
$n=-4, \ldots 12$. The ground state energy is
$E_g = -4{\cal J}_1/3$, corresponding to the 90-fold degenerate pure spin ice states (the
configurations in which there are two spins pointing into and two spins pointing out of each of the eight tetrahedra in our system).
Fig.\ \ref{j2_all_1}(a) shows the result for ${\cal J}_{3,4}= 0 $
and ${\cal J}_2 \approx 0$; here the eigenvalues are almost discrete. A continuous spectrum emerges 
when any of the constants ${\cal J}_{2,3,4}$ become non-zero. 

\begin{figure}[ht]
\centering
\subfloat[${\cal J}_{3}=-0.05$ ]{
  \includegraphics[width=40mm]{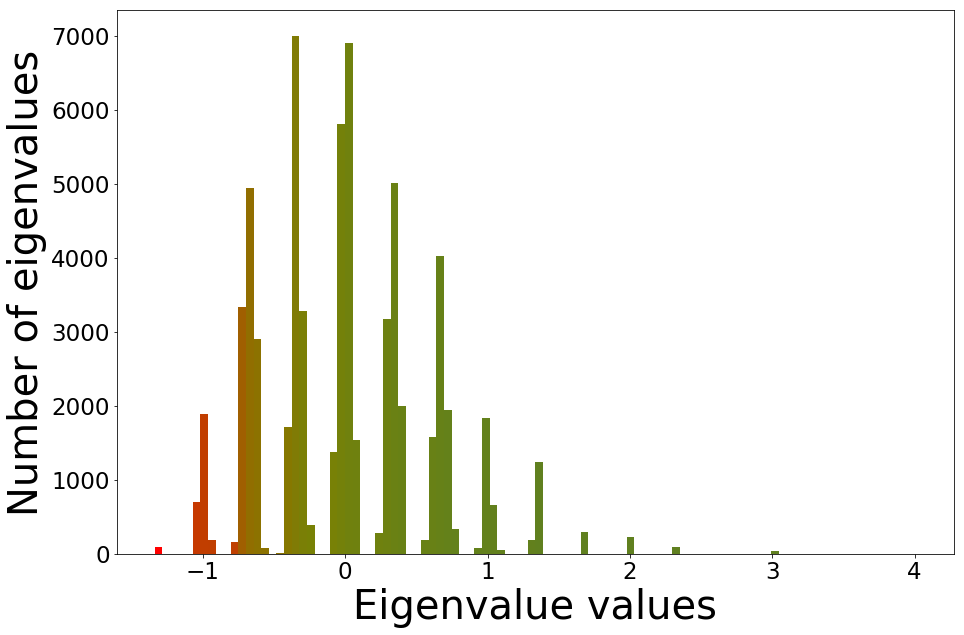}
}
\subfloat[${\cal J}_{3}=-0.2$]{
  \includegraphics[width=40mm]{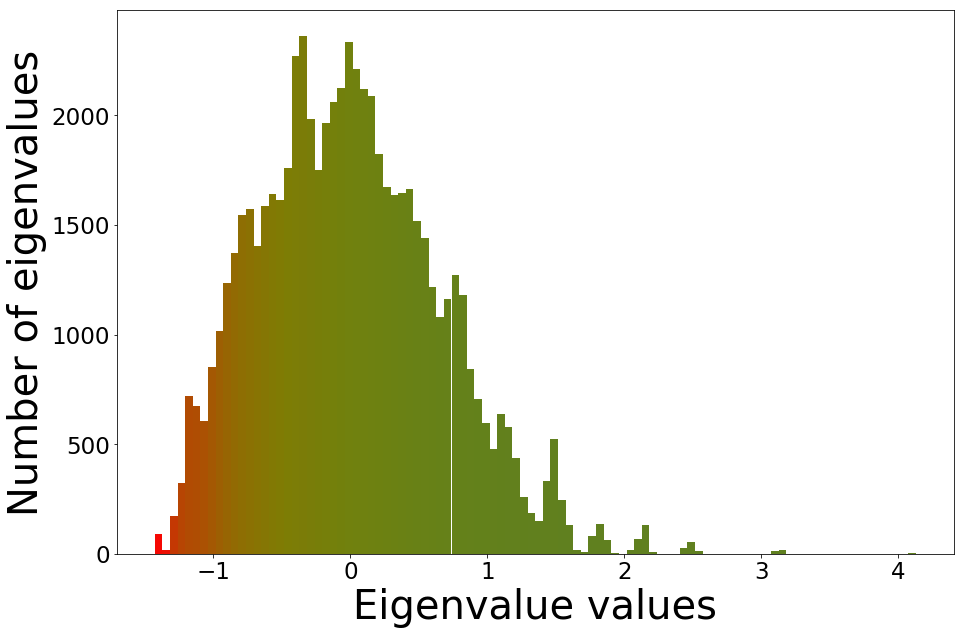}
}
\hspace{0mm}
\subfloat[${\cal J}_{3}=-0.35$ ]{
  \includegraphics[width=40mm]{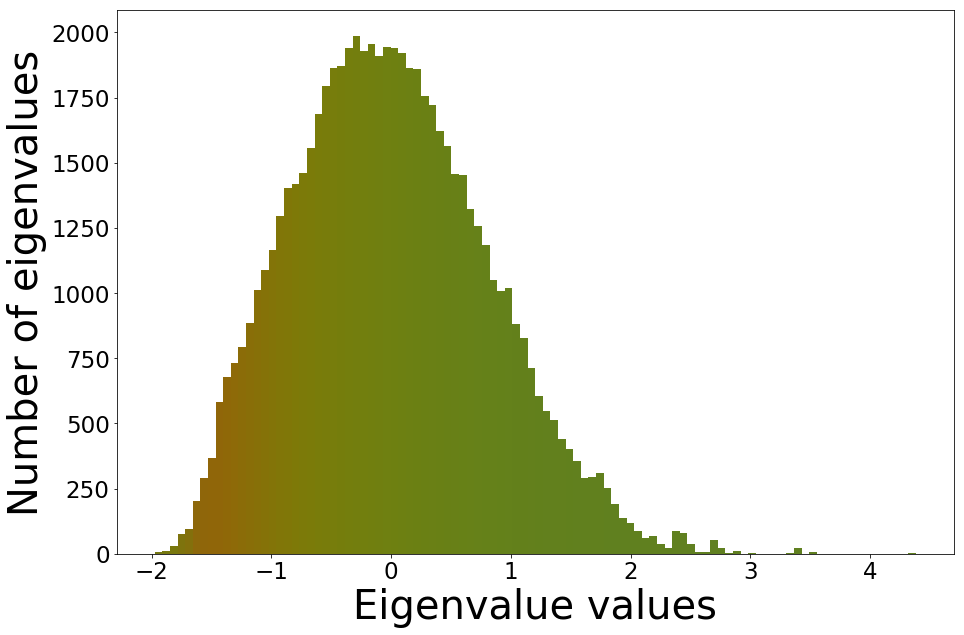}
}
\subfloat[${\cal J}_{3}=-0.5$]{
  \includegraphics[width=40mm]{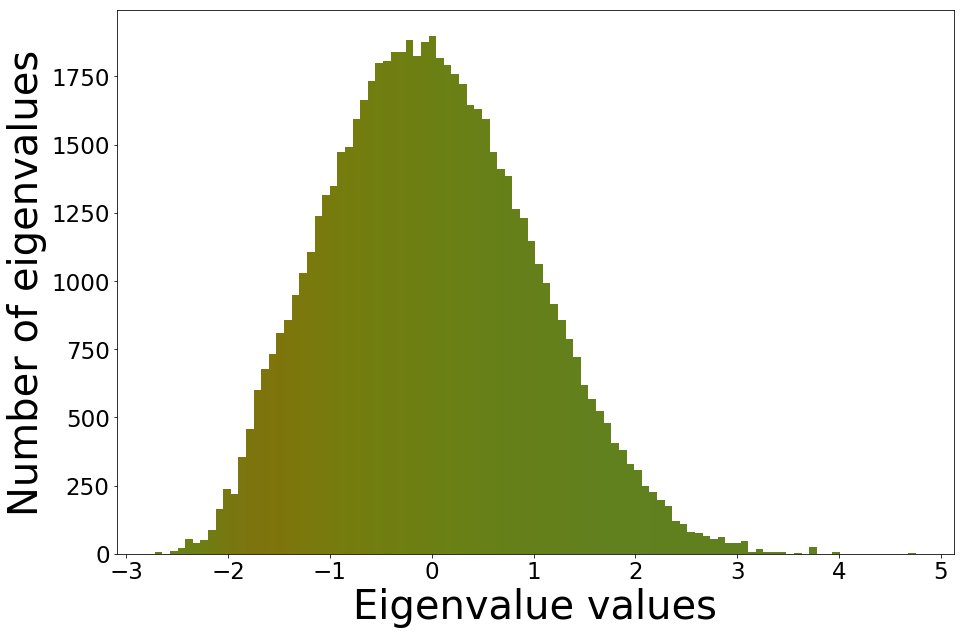}
}

\caption{Spectrum of $H$ for ${\cal J}_1 = -1$, ${\cal J}_2 = {\cal J}_4 = 0$,  and  different  values of ${\cal J}_{3}$.
\label{j3_all}}
\end{figure}

\begin{figure}[ht]
\centering
\subfloat[${\cal J}_{4}=-0.1$ ]{
  \includegraphics[width=40mm]{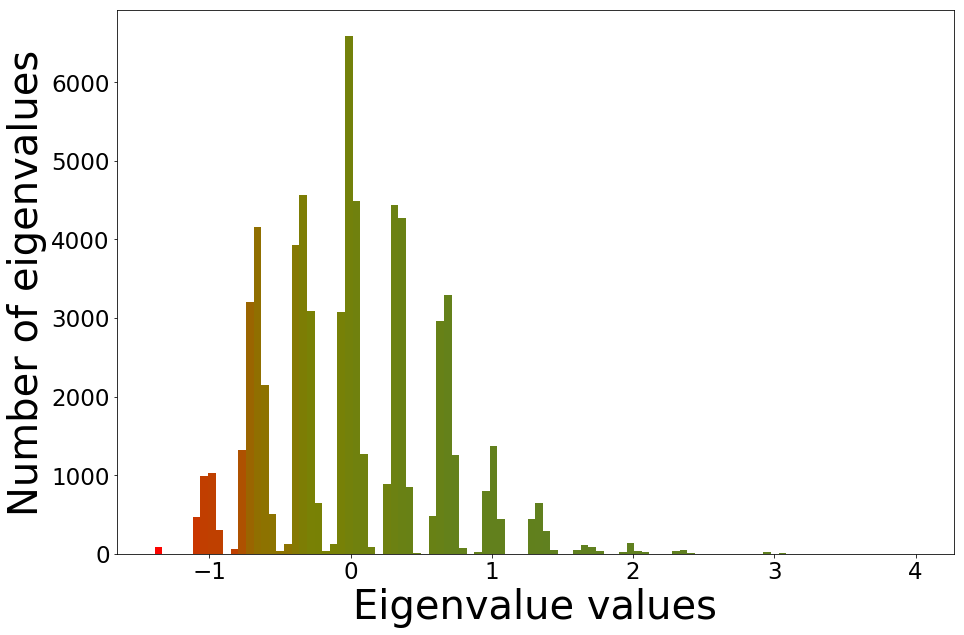}
}
\subfloat[${\cal J}_{4}=-0.4$]{
  \includegraphics[width=40mm]{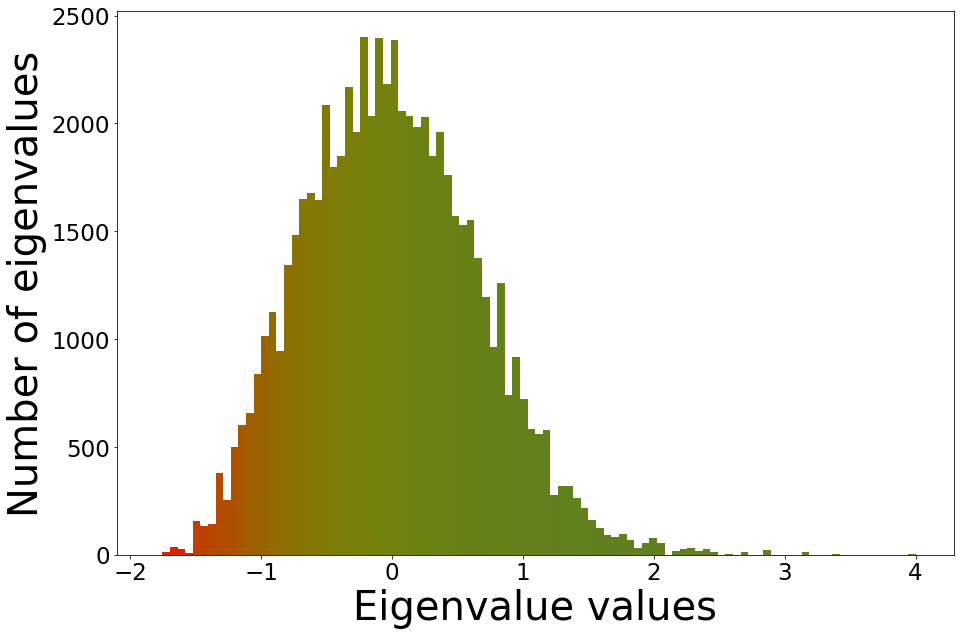}
}
\hspace{0mm}
\subfloat[${\cal J}_{4}=-0.7$ ]{
  \includegraphics[width=40mm]{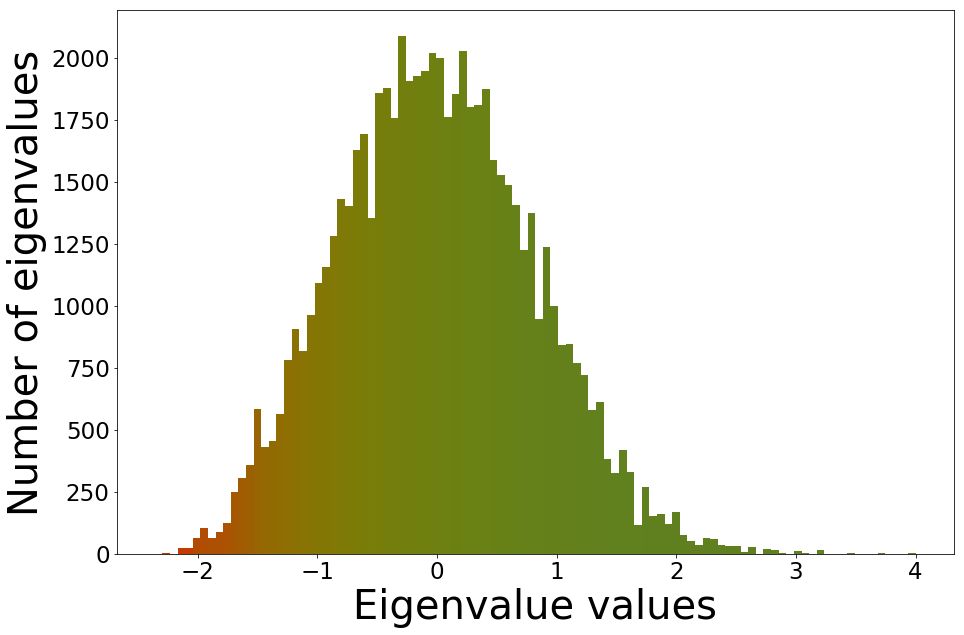}
}
\subfloat[${\cal J}_{4}=-1$]{
  \includegraphics[width=40mm]{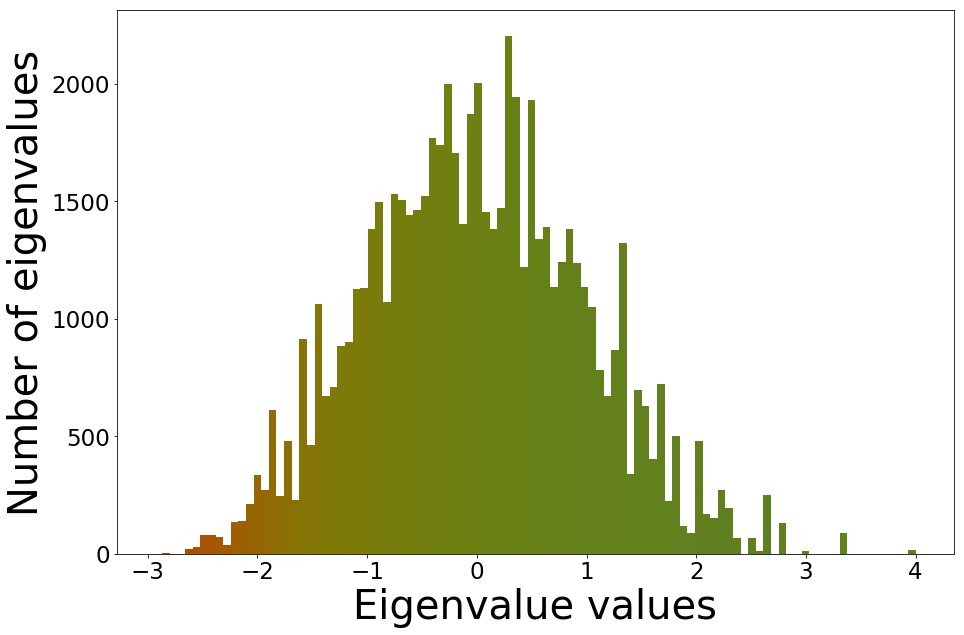}
}

\caption{Spectrum of $H$ for ${\cal J}_1 = -1$, ${\cal J}_2 = {\cal J}_3 = 0$,  and  different  values of ${\cal J}_{4}$.
\label{j4n_all}}
\end{figure}

\begin{figure}[ht]
\centering
\subfloat[${\cal J}_3 ={\cal J}_{4}=-0.05$ ]{
  \includegraphics[width=40mm]{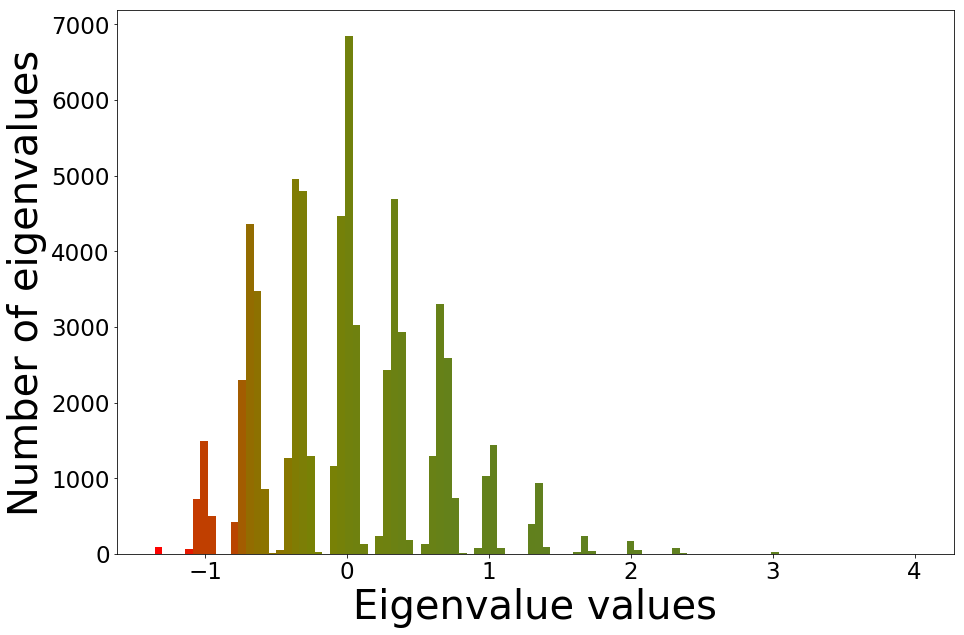}
}
\subfloat[${\cal J}_3 ={\cal J}_{4}=-0.2$]{
  \includegraphics[width=40mm]{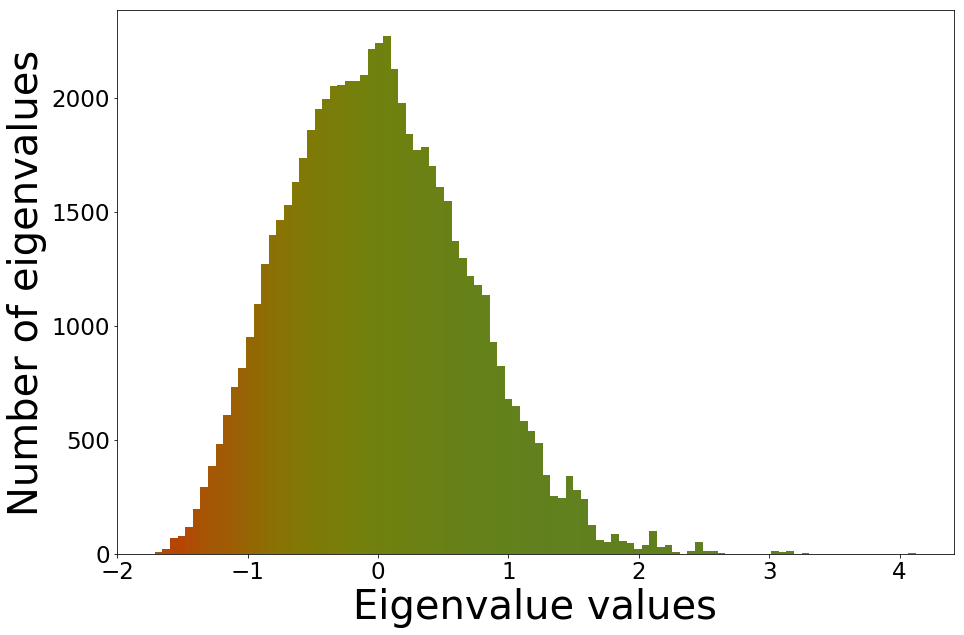}
}
\hspace{0mm}
\subfloat[${\cal J}_3 ={\cal J}_{4}=-0.35$ ]{
  \includegraphics[width=40mm]{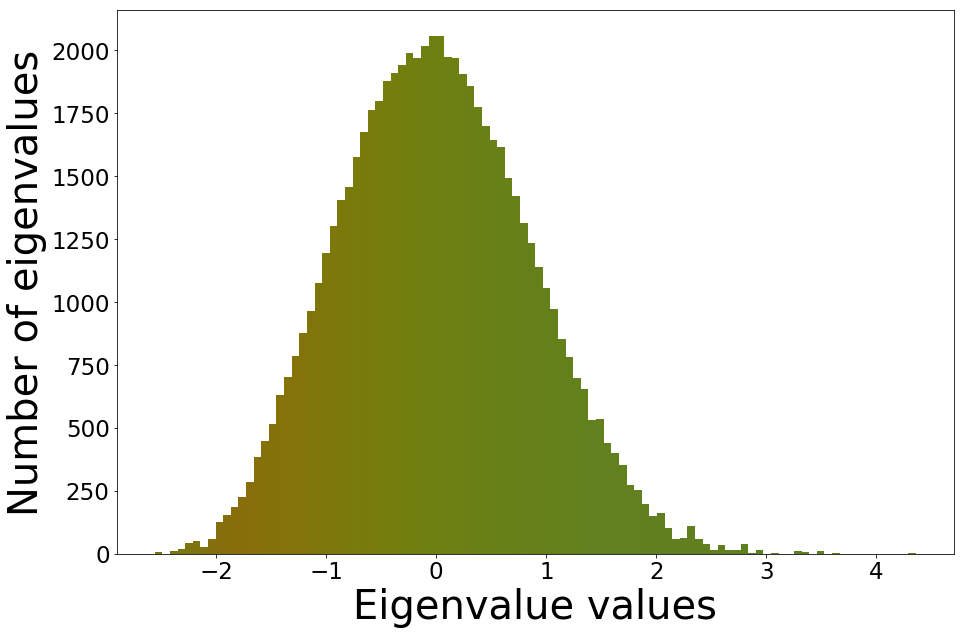}
}
\subfloat[${\cal J}_3 ={\cal J}_{4}=-0.5$]{
  \includegraphics[width=40mm]{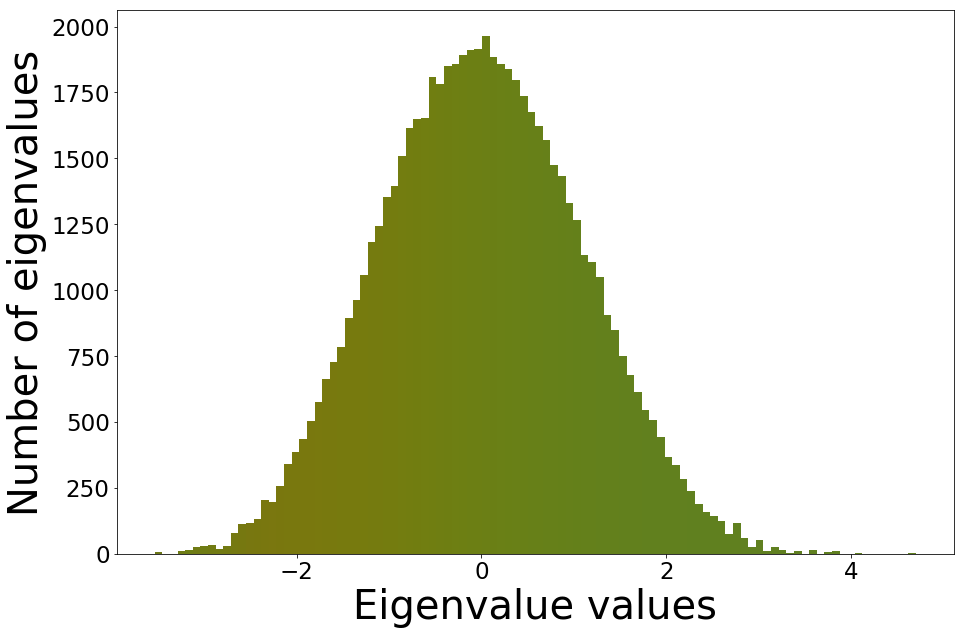}
}
\caption{Spectrum of $H$ for ${\cal J}_1 = -1$, ${\cal J}_2 =  0$,  and  different  values of ${\cal J}_3 ={\cal J}_{4}$.
\label{j4n_all}}
\end{figure}

Although the block-diagonalization approach followed here is a well-established, mathematical one, a distinctly physical picture emerges from the construction.  Each basis ket $|u_i\rangle = |\pm\pm\pm \ldots\rangle$ corresponds to a classical  state in which each of the spins points into or out of the tetrahedra in the lattice. States with exactly two spins pointing into and two spins pointing out of every tetrahedron are known as ``spin ice states. Other possible configurations 
are those with three spins pointing into and one spin pointing out of (or vice versa) and all spins pointing into or out of a tetrahedron.  All configurations can be classified by a set of numbers, $\{n_1,n_2,n_3\}$,
where $n_1$ is the number of 2-in-2-out tetrahedra,
$n=2$ is the number of 3-in-1-out/ 1-in-3-out tetrahedra, and $n_3$ is the number of all-in/all-out 
tetrahedra, where
$\sum_i n_i$ equals the total number of tetrahedra, 
which is $N/2$ for periodic boundary conditions.

The symmetrized kets $|\phi_i\rangle$,
which are the basis for the block-diagonal representation of $H$, are linear combinations of the basis kets. The $|\phi_i\rangle$  are found by applying all of the symmetry elements to one basis ket. 
Since the application of any symmetry element to a basis ket will rearrange the spins in the ket {\em while preserving the numbers} $n_i$, each symmetrized ket can be characterized by the same set of numbers $n_i$.  In particular, by construction, there will be a set of symmetrized
states that are pure spin ice states with $n_1 = 8$ and $n_{2,3} = 0$. 

The colour scheme of Figs.\ \ref{j2_all}-\ref{j4n_all} represents the average values of the $n_i$ in each bin of the histogram, using red for $n_1$, green for $n_2$ and blue for $n_3$. The colour green dominates these plots because the 3-in-1-out/1-in-3-out configurations are more numerous. $n_1$ (the number of 2-in-2-out tetrahedra) is largest at the lowest energy part of the spectrum and appears as bright red.  Small variations of the coupling constants produces perturbative shifts of their eigenvalues and mixing of the eigenstates which smears the colours toward a uniform green. 

When ${\cal J}_{2,3,4}=0$ the basis states $|u_i\rangle = |\pm\pm\pm \ldots\rangle$ 
are eigenstates of $H$ with energy $\frac{{\cal J}_1}{6}(- n_1 + 3 n_3)$. 
All of the states are at least two-fold degenerate, with the highest degeneracy associated with states containing 3-in-1-out/1-in-3-out 
configurations.  
The symmetrized kets $|\phi_i\rangle$,  which are the basis of the block-diagonalized Hamiltonian, are linear combinations of degenerate states
and are the eigenstates of the general Hamiltonian (\ref{eq:hamil}) to zeroth
order in degenerate perturbation theory.  By construction, these states are highly entangled, although more involved calculations that account for the degeneracies in the representations, as well as the near-degeneracies in the spectrum for small values of the coupling constants
${\cal J}_{2,3,4}$ (which become relevant at non-zero temperature) are needed to properly characterize the entanglement of the system. 

To summarize, in this paper we discuss the application of group theory methods to block-diagonalize the Hamiltonian matrix of a 16-site pyrochlore magnet. The method is essentially analytic, such that in principle the unitary matrix that block-diagonalizes $H$ can be determined with absolute precision, however in our implementation this is handled numerically. Once calculated, the unitary transformation need only be applied once to each of the four terms in $H$; we then vary the coupling constants and find {\em all} the eigenvalues numerically.  In this way, one can study the phase diagram spanned by the coupling constants of the model \cite{chen_curnoe_2023}.



\begin{acknowledgments}
We thank Oliver Stueker for assistance with
using the resources at Compute Canada and
Kyle Hall for helpful discussions about coding. This work was performed using the resources at Compute Canada and supported by the Natural Sciences and Engineering Research Council of Canada.
	\end{acknowledgments}

	\appendix


\section{The space group}

In pyrochlore crystal, with space group $Fd\bar{3}m$, the
magnetic sites are located at the 16d Wyckoff position, which has site symmetry $D_{3d}$.   This site symmetry plays an important role in quantum magnets, because it ensures that
the lowest energy state of the 
electronic part of the magnetic ion is either a singlet or a doublet.  In many real crystals the magnetic ion is a rare earth,  and 
the electronic ground state is a doublet that is well-separated
from higher energy levels; moreover, in some cases the doublet has exactly the same symmetry as a spin-1/2 spinor, which allows one to model these systems as spin-1/2 states residing at the 16d positions, which are the vertices of a network of corner-sharing tetrahedra - the quintessential 
frustrated quantum magnet. 

The point group $D_{3d}$ contains a 3-fold axis (see Fig.\ \ref{fig:pyro}), which is the axis of highest symmetry. It is convenient to express the electronic total angular momentum with respect to that 
axis; also, by construction, angular momentum states $|j,m\rangle$ are eigenstates of $J_z$, 
thus to each magnetic site we attach 
a set of local axes, such
that the local $z$-axis points in the direction of the 3-fold axis
of $D_{3d}$ (see Ref.\ \cite{Curnoe2007_PhysRevB_2007} for more details). 

When a space group operation is applied on a crystal with periodic boundary conditions, equivalent sites (such as the 16d positions) will be permuted. The permutation can be found by explicit application of the space group operation on the position of the magnetic sites. In addition, the local axes attached to each site may also be rotated by one of the elements of $D_{3d}$. The elements of $D_{3d}$ are 
$\{E,C_{3z},C_{3z}^{2}\} \times \{E,C_{2y}\}
\times \{E , I\}$.
The actions of these 
operators on the spin states are:
$C_{3z}|\pm\rangle 
= \exp\left(\mp i \frac{\pi}{3}\right)|\pm\rangle$,
$C_{2y}|\pm \rangle=
\mp|\mp\rangle$
and $I|\pm\rangle = (-)|\pm\rangle$. The element $I$ may produce a sign change depending on the parity of the electronic state, but since the phase factors for all sites will be multiplied, and since we will always consider an even number of spins, this factor is of no relevance.  
For each space group element we find and store the permutation of the sites it produces and the action on the local sites (one of the 12 elements of $D_{3d}$).

The sub-group 
 $\{O_h\}\times \{E,\tau_1, \tau_2, \tau_3\}$ corresponds to a system with 16
magnetic sites arranged inside a cube, with periodic boundary conditions assumed.  
The group elements 
 are  listed in the first row of Table \ref{16site-char}. 
Here $C_{n}$ is an $n$-fold rotation, $I$ is inversion and $\tau$ is a 
translation. Some of the rotations are screw rotations, which are rotations followed by a translation along the rotation axis. Similarly, some of the improper rotations (reflections) are actually glides, which are 
reflections followed by a translation within the mirror plane.  The details of all the group operations can be found in Ref.\ 
\cite{international_TableA}. When applied to the spins states of a cubic cell, the group operations will permute the 
sites, possibly introduce a phase, and possibly reverse the spin at each site. 

The different columns
in Table \ref{16site-char}  refer to different {\em classes} of group elements, whereby
 two elements ${\cal A}$ and ${\cal B}$ belong to the same class if there is a group element ${\cal G}$ such that 
 ${\cal B} = {\cal G} {\cal A} {\cal G}^{-1}$. Generally speaking, elements in the same class are physically similar to each other, and the matrix representations of elements belonging to the same class have the same traces.  The rows 
 refer to different 
 irreducible representations $\Gamma^{(i)}$ of the group; their dimensions $d_i$ are given in the second column. The set of numbers in each row of the table are the character of the representation (the
traces of the matrix representations of the group elements).
 Representations are equivalent if they have the same character; the character of a reducible representation is the sum of the characters of the IRs in its decomposition.

\section{Block Diagonalization}



The unitary transformation that block-diagonalizes the Hamiltonian is found by generating the complete set of orthogonal, normalized, symmetrized kets  belonging to each IR by 
applying the projection operators ${\cal P}_{\kappa\kappa}^{(j)}$ to the basis kets, as follows. 
For a given basis ket 
$|\pm\pm\pm \ldots\rangle$
we first generate all of its partners by applying to it each of the symmetry elements $P_R$. The result will be a basis ket (which may be the given basis ket we started with) with a phase (which may be 1). A set of $d$ unique partners is the basis of a $d$-dimensional reducible representation; its decomposition into IR's is then determined using characters. 
In our case, because the group elements transform single kets into single kets, the matrices representing the group elements will have only one non-zero element (a phase) in each row and column, and the trace will simply be the sum of the phases attached 
to the kets which transform into themselves. 
Each one-dimensional IR will appear at most once in this decomposition, but the multi-dimensional IR's may appear more than once.
For example, the 16-site ket
$|++++++++++++++--\rangle$
is one of 96 partners that are a basis for the reducible
representation
$A_{1g}\oplus A_{2g} \oplus
A_{1u} \oplus A_{2u} + 2 E_g$.  In our 16-site system, the smallest set of partners contains two basis kets, 
$|+++\ldots +\rangle$
and $|--- \ldots - \rangle$, which are a basis for $E_g$, while the largest sets of partners contain 192 partners (the number of group elements), which are a basis for the {\em regular} representation, 
$\sum d_i \Gamma^{(i)}$.

Once the set of partners and the decomposition of its representation has been determined, we apply the projection operators 
${\cal P}_{\kappa\kappa}^{(j)}$ to the $d$ partners in order to find the 
symmetrized basis kets. 
The symmetrized basis kets
belonging to the 1D IR's are easily obtained: since
they each contain all of the partners, any of the partners can be used to generate them by one 
application of ${\cal P}_{11}^{(j)}$. The difficulty with the multi-dimensional representations is that their symmetrized kets do not necessarily contain all of the partners, and so it may take many applications of ${\cal P}_{\kappa\kappa}^{(j)}$
on different partners in order to find all of the symmetrized kets.

To minimize the number of failed applications of ${\cal P}_{\kappa\kappa}^{(j)}$ to a set of partners, 
for each 
representation $j$ we construct a $d\times d_j$ `{\em flag}' array, where $d$ is the number of partners and $d_j$ is the dimension of the $j$th representation.  The rows are indexed by the partner number $p=1$ to $d$ and the rows by the block number $\kappa = 1$ to $d_j$. This array keeps track of the number of times a partner appears in the symmetrized kets belonging to each block of the $j$th IR, which can be at most one. We observe that the sum of the entries in a row is the same for all rows. 
Thus we generate the 
symmetrized kets for the $j$th representation as follows. {\em i)} 
Apply to the first partner $p=1$ the projectors ${\cal P}_{\kappa\kappa}^{(j)}$ for 
$\kappa = 1$ to $d_j$. 
The non-zero results are symmetrized kets belonging to the $\kappa$th block of the $j$th IR. 
{\em ii)} For each partner $p$ appearing in a symmetrized ket, change
the entry $(p,\kappa)$ in the flag matrix
to one, where $p$ is the partner number. Thus 
the first row of the flag array is determined, and 
the sum of its elements (which we will call `{\em flag-sum}'
can be found. 
{\em iii)} While the sum of 
the elements in the second row of the flag array is less than {\em flag-sum},
apply to the second partner $p=2$ the projectors ${\cal P}_{\kappa\kappa}^{(j)}$ for the
values of $\kappa$
where the $(2,\kappa)$ entries of the flag array are zero.  A non-zero 
result is a symmetrized ket; 
for each partner appearing in a symmetrized ket, change
the entry $(p,\kappa)$ in the flag matrix
to one, where $p$ is the partner number.
{\em iv)} Consider 
subsequent partners $p$ until 
$d$ symmetric basis vectors have been found.

We also create a flag array of length $2^{16}$
which keeps track of original basis kets.  Each time a set of partners (of original basis kets) is found and all the 
symmetrized kets have been generated from that set of partners then all the partners are flagged.  Thus
the procedure for finding all the symmetrized kets is as follows.  
{\em i)} Beginning
with the first ket, 
$|+++\ldots +\rangle$, find
its partners and
the symmetrized kets as described above; flag all of the partners. 
{\em ii)} Repeat
for all kets  numbered $n=2$ to $2^{N}$, skipping any kets 
that have been flagged.


\section{Extension to 32 sites}

%



\begin{widetext}
\begin{center}
\begin{table}[h]
\begin{tabular}{|r|r|r|r|r|r|r|r|r|r|r|r|r|r|r|r|r|r|r|r|r|}
\hline
&$E$&$12C_2$&$32C_3$&$32C_3$&$24C_2'$&$48C_4$&$4I$&$24IC_2$&$32IC_3$&$32IC_3$&$24IC_2'$&$12IC_2'$&$48IC_4$&$6\tau $&$12\tau C_2$&$12TC_2'$&$12\tau C_2'$&$12\tau IC_2'$&$\tau_4$&$4I\tau_4$ 
    \\ \hline
$A_{1g}$ &1&1 &1 &1 &1 &1 &1 &1 &1 &1 &1 &1 &1 &1 &1 &1 &1 &1 &1 &1   \\ \hline
$A_{2g}$&1&1 &1 &1 &-1&-1&1 &1 &1 &1 &-1&-1&-1&1 &1 &-1&-1&-1&1 &1   \\ \hline
$E_{g}$&2&2 &-1&-1&0 &0 &2 &2 &-1&-1&0 &0 &0 &2 &2 &0 &0 &0 &2 &2     \\ \hline
$T_{1g}$&3&-1&0 &0 &-1&1 &3 &-1&0 &0 &-1&-1&1 &3 &-1&-1&-1&-1&3 &3     \\ \hline
$T_{2g}$&3&-1&0 &0 &1 &-1&3 &-1&0 &0 &1 &1 &-1&3 &-1&1 &1 &1 &3 &3    \\ \hline
$A_{1u}$&1&1 &1 &1 &1 &1 &-1&-1&-1&-1&-1&-1&-1&1 &1 &1 &1 &-1&1 &-1   \\ \hline
$A_{2u}$&1&1 &1 &1 &-1&-1&-1&-1&-1&-1&1 &1 &1 &1 &1 &-1&-1&1 &1 &-1     \\ \hline
$E_{u}$&2&2 &-1&-1&0 &0 &-2&-2&1 &1 &0 &0 &0 &2 &2 &0 &0 &0 &2 &-2     \\ \hline
$T_{1u}$&3&-1&0 &0 &-1&1 &-3&1 &0 &0 &1 &1 &-1&3 &-1&-1&-1&1 &3 &-3     \\ \hline
$T_{2u}$&3&-1&0 &0 &1 &-1&-3&1 &0 &0 &-1&-1&1 &3 &-1&1 &1 &-1&3 &-3    \\ \hline
$X_{1}$      &6&2 &0 &0 &2 &0 &0 &0 &0 &0 &0 &0 &0 &-2&-2&-2&-2&0 &6 &0        \\ \hline
$X_{2}$      &6&2 &0 &0 &-2&0 &0 &0 &0 &0 &0 &0 &0 &-2&-2&2 &2 &0 &6 &0           \\\hline
$X_{3}$      &6&-2&0 &0 &0 &0 &0 &0 &0 &0 &-2&2 &0 &-2&2 &0 &0 &2 &6 &0         \\ \hline
$X_{4}$      &6&-2&0 &0 &0 &0 &0 &0 &0 &0 &2 &-2&0 &-2&2 &0 &0 &-2&6 &0          \\ \hline
$L_{1}$      &4&0 &1 &-1&0 &0 &-2&0 &1 &-1&0 &2 &0 &0 &0 &2 &-2&-2&-4&2           \\\hline
$L_{2}$      &4&0 &1 &-1&0 &0 &2 &0 &-1&1 &0 &-2&0 &0 &0 &2 &-2&2 &-4&-2         \\ \hline
$L_{3}$      &4&0 &1 &-1&0 &0 &-2&0 &1 &-1&0 &-2&0 &0 &0 &-2&2 &2 &-4&2          \\ \hline
$L_{4}$      &4&0 &1 &-1&0 &0 &2 &0 &-1&1 &0 &2 &0 &0 &0 &-2&2 &-2&-4&-2         \\ \hline
$L_{5}$      &8&0 &-1&1 &0 &0 &-4&0 &-1&1 &0 &0 &0 &0 &0 &0 &0 &0 &-8&4          \\ \hline
$L_{6}$      &8&0 &-1&1 &0 &0 &4 &0 &1 &-1&0 &0 &0 &0 &0 &0 &0 &0 &-8&-4         \\ \hline
\end{tabular}
\caption{Character table for the symmetry group of the $32$-site system.\label{32char}}
\end{table} 
\end{center}
\end{widetext}

The 32-site system contains 8 tetrahedra
within two adjoining conventional (cubic) cells. 
The symmetry group is 
$\{O_h\} \times \{E, \tau_1, \ldots \tau_7\}$, where the additional translations are : $\tau_4=(1,0,0)$, $\tau_5=\tau_1 + \tau_4 = (1,1/2,1/2)$, $\tau_6 = \tau_2 + \tau_4 =(3/2,0,1/2)$ and $\tau_7= \tau_3 + \tau_4 = (3/2,1/2,0)$.  The symmetry group contains 384 elements divided into 20 classes; its character table is given in Table \ref{32char}. The decomposition of the representation generated by the $2^{32}$ basis kets is given in Table
\ref{32blocks}.

\begin{table}[h]
\begin{tabular}{c|c|r} 
\hline 
IR & dimension & block size 
\\
\hline 
$A_{1g}$ & 1& 11,201,728\\ 
$A_{2g}$ & 1&11,197,500 \\
$A_{1u}$& 1 & 11,174,256 \\
$A_{2u}$ & 1& 11,178,412 \\
$E_{g}$ & 2& 22,398,924\\
$E_{u}$ & 2& 22,352,476\\
$T_{1g}$& 3& 33,588,204 \\
$T_{2g}$& 3& 33,592,296\\
$T_{1u}$ & 3& 33,522,668\\
$T_{2u}$ & 3& 33,518,632 \\
$X_{1}$ & 6& 67,102,708 \\
$X_{2}$& 6&   67,102,772 \\
$X_{3}$ & 6& 67,114,964\\
$X_{4}$ & 6& 67,106,836 \\
$L_{1}$ & 4 & 44,755,688\\
$L_{2}$ & 4 & 44,722,968\\
$L_{3}$ & 4 & 44,763,880\\
$L_{4}$ & 4 & 44,714,776\\
$L_{5}$ & 8 & 89,519,376\\
$L_{6}$ & 8& 89,437,408 \\
\hline 
\end{tabular}
   \caption{The decomposition of the representation generated by the $2^{32}$ basis kets of two adjacent cubes containing a total of 32 sites. 
    The first column lists the IR's of the 
   symmetry group, the second column lists their dimension (degeneracy), and the final column gives the size of each block (the number of copies $a_i$ of each IR).
    \label{32blocks}}
\end{table}

\bibliographystyle{apsrev4-1}
    \bibliography{main}

\end{document}